\documentclass{aa}  

\pdfoutput=1
\usepackage[figuresright]{rotating}
\usepackage{graphicx}
\usepackage{natbib}
\bibpunct{(}{)}{;}{a}{}{,}
\usepackage{txfonts}

\begin{document}
   \title{Mapping  the interface between the Local and Loop I bubbles \\ using Str\"omgren photometry }

   \author{W. Reis
          \and
          W.J.B. Corradi
          }

   \offprints{W. Reis}

   \institute{Departamento de F\'isica - ICEx - UFMG, Caixa Postal 702, 30.123-970 Belo Horizote - MG, Brazil\\
              \email{wilsonr@fisica.ufmg.br} \\
              \email{wag@fisica.ufmg.br}
               }

   \date{Received May  24, 2007; accepted May 24, 2007}

 
  \abstract
   {The Sun is located inside an extremely low density region   of  quite irregular shape called the
   Local Bubble or Local Cavity. The fraction of this cavity filled with extremely hot gas is 
   known as the Local Hot Bubble. Close to the Local Bubble, there is an even larger cavity known 
   as Loop I. A ring-like feature observed in X-ray and H{\sc I}
   has been proposed as the contour of the bubbles interaction zone around 70 pc.
   }
   {Our goal is to identify  the interface
between the Local  and Loop I Bubbles and discuss the ring's existence  
using Str\"omgren $uvby$H$\beta$ data.
   }
   {We have used the $uvby$H$\beta$ data of the General
Catalogue of Photometric Data,
 covering the region defined by the Galactic coordinates:
$250\degr \leq l \leq 50\degr$ and $-60\degr \leq b \leq 60\degr$
to obtain $E(b-y)$ colour excess and distances.
A set of exclusion criteria have been applied to eliminate the values inappropriate
to the study of the interstellar reddening distribution.
The final sample is composed of 4346 stars located  up to 500 pc from the Sun.
   }
  { {
 The expected transition to $E(b-y) \approx  0\fm070 - 0\fm100$, corresponding to the ring's column density,
 occurs on the western  part of the ring at $d=110 \pm 20$ pc, whereas on the eastern side it is not clearly
 seen before $d=280 \pm 50$ pc. Near the Galactic plane the presence of the dark clouds is clearly established 
 by $E(b-y) \geq 0\fm100$ in the western side at about 100 -- 150 pc and about 120 -- 180 pc in the eastern side.
Beyond these distance ranges the number of unreddened stars decreases considerably indicating the location of these
large dark cloud complexes.
In the southernmost part of the ring the reddening remains very low, typically  $E(b-y) = 0\fm020$ in all
 its extension, except towards the Mensa constellation where a possible transition is observed at $d = 200 \pm 20$ pc.
In the northernmost part the colour excess increases with  distance in a gradual way,
such that $0\fm020 \leq E(b-y) \leq 0\fm040$ becomes predominant only after $d = 120\pm$15  pc.
}}
   {If the ring really exists the colour excess distribution indicates that it is very fragmented and
distorted. However,  the different characteristics of the reddening inside and along the
ring do not support the  existence of a ring.}


   \maketitle
%

\section{Introduction}

The Sun is located inside an extremely
 low density region ($n_{HI} \leq 0.005$ cm$^{-3}$)  of  quite irregular shape called the
 Local Bubble (LB) or Local Cavity.
The fraction of this cavity 
  filled with extremely hot gas ($T \approx 10^{6}$ K)
 is known as the Local Hot  Bubble  (LHB)
 \citep[e.g.,][]{paresce84,cox87,snowden90,warwick93,leroy99,sfeir99,breit2000,lallement2003,lallement2005}.

In the direction of the  Scorpio-Centaurus OB  association  (Sco-Cen), 
there is an even larger cavity known as 
   Loop I.
It is believed that  Loop I has been formed by the  action of the stellar wind and
supernova explosions of the stars located in the Sco-Cen,  that
acting on the interstellar material, left after the star formation,  created a
 bubble of gas and dust concentrical to this association.
  \citep[e.g.,][]{weaver79,iwan80,degeus89}.

 Historically, the concept of the
formation of the LB is a dispute among three views: (I) the view that one or more supernova explosions,
near  the Sun, formed the cavity and the soft X-ray emitting region
\citep[e.g.,][]{smith74,cox82,cox87,smith01,apellaniz01,breit02,fuchs2006};
 (II) the view that
 the Loop I superbubble has been formed by  successive epochs of star formation in the Sco-Cen, having expanded
asymmetrically into the low density inter-arm region surrounding the Sun
 \citep[e.g.,][]{frisch81,frisch83,bochkarev87,frisch95,wolleben2007}. (III) some alternative views  
where the notion of the low density region around the Sun as a bubble is dismissed \citep[e.g.,][]{bruh96,mebold98,lepine03}

Under class ``I" models, \citet{smith74}  have shown that the number of supernova in the Galaxy  
 has been high enough for 
 supernova remnants (SNRs) to  occupy a significant fraction of the interstellar medium.

A more detailed model  for the Local Bubble was developed by \citet{cox82},
 in which they  considered that only one  supernova occurring in the vicinity of the Sun
 in an environment of low density ($ n \approx 0.004$ cm$^{- 3}$)  would have  been sufficient to create what we observe today.
However, if  the surrounding interstellar density is about  0.1 cm$^{- 3}$,  a
 sequence  of supernova explosions  in the same region would be necessary to construct the cavity of
 low density of the observed size  and to  reheat its interior.
 Models with this line of thought have been developed by \citet{cox86,cox87,cox93} and \citet{smith97}.

\citet{smith01} have
shown that two or three supernova explosions in a diffuse interstellar medium would be able 
to generate a hot bubble like the LB. In this model the supernovae explosions would be
random.
Following the same line of reasoning,
  \citet{apellaniz01} suggested that the LB and Loop I  could have 
started as a single entity, originating from supernova explosions in the  
 Upper-Centaurus-Lupus (UCL) sub-group  of the Sco-Cen. Later, three supernovae explosions in the 
Lower-Centaurus-Crux (LCC) subgroup of the Sco-Cen would have been responsible for the expansion in 
the direction of the LB.

\citet{breit2000} show a schematical representation
of the interaction between the LB and the neighboring Bubble Loop I \citep[see][~Fig.~1]{breit2000}.
Using Hipparcos stellar distances and the results of the kinematical analysis suggested by
\citet{asiain}, \citet{breit02} suggested that  thirteen million years ago, the center of mass
 of the Pleiades  B1 subgroup  was located inside the volume of the LB,
and  10 - 20 supernovae explosions  in this subgroup would have been  responsible  
for the LB creation.

Assuming a background medium pre-structured by previous generations of supernova explosions,
\citet{breit06} have suggested an age of $14.5^{+0.7}_{-0.4}$ Myr for the LB, which would have been
reheated by 19 supernovae explosions to date. \citet{fuchs2006}, instead, suggested
that  14 - 20 supernovae explosions originating in LCC and UCL would have been responsible for
the LB origin.  

Under the notion of class ``II" models, \citet{frisch81,frisch83,frisch86,frisch95} proposed that
the LB is part of a superbubble, created by the stellar  activity  in the Sco-Cen, 
that expanded through the  low density region among  the Galaxy spiral arms. 
\citet{frisch98} shows a schematical representation of the three shells from the star 
 formation in the Sco-Cen \citep[see][~Fig.~2]{frisch98}.

\citet{bochkarev87} suggested that the solar system is near the limit of a  low density region
 $(n \approx 1 \times 10^{-3} - 4 \times 10^{-3}$ cm$^{-3})$ filled with hot gas $(10^{6}$ K)
 called the Local Cavern. The center of the Local Cavern would be located in the Sco-Cen region  and would be involved
 by a gas envelope that would be  observed as H{\sc I} filaments  with a mass of the order of $10^{6} M_{\odot}$,
  the distance between the center and the envelope  being  approximately 150-200 pc.  

  \citet{wolleben2007} proposed a model consisting of two synchroton-emitting shells, namely S1 (LCC bubble)
and S2 (Loop I bubble), expanding with different velocities. In this picture S1 is about 6 Myr old and S2 is 1-2 Myr old.
The Sun would reside within S1, but near its border. The shock front of S2 would have hit S1 just recently 
($10^{4}$ yr ago or less), giving rise to the X-ray emission observed as the North Polar Spur. 

Under the alternative class ``III" models \citet{bruh96} and \citet{mebold98}
argue that a Bubble may not even exist, the LB being only the intersecting contour
of the neighbouring bubbles. Another idea has been proposed by \citet{lepine03}
where the presence of sheet-like structures of gas and dust near the Sun would be created by  the
shock of a spiral arm with the interstellar medium.

Due to the proximity of the LB and Loop I, it is believed that they may be 
interacting. Indications of the existence of a ring-like  structure of 
dense neutral matter that  would be the contour of the interaction zone 
between the two bubbles has been identified in X-ray and neutral hydrogen 
data by \citet{egger95}. Such an interface would be located in the region 
defined by the Galactic coordinates: $275^{o} \leq l \leq 35^{o}
 \ \mathrm{and} \  -45^{o} \leq b \leq 45^{o}$. It has been also suggested 
by  \citet{frisch2007} that the interstellar large scale structures, identified 
in her mean extinction map in this direction, would make up the ring (see her Fig. 1). 

Different authors present different values for the distance of this interaction 
zone between the LB and Loop I: Centurion \& Vladilo (1991) analyzing UV
spectra  of eight stars in the region defined by the Galactic coordinates
$ 315\degr \leq l \leq 330\degr \ \mathrm{and} \  15\degr \leq b \leq 25\degr$
suggested that the gas wall is located at a distance of 40 $\pm$ 25 pc from the Sun;
\citet{egger95} using data compiled  by \citet{fruscione}  suggested that the 
distance would be approximately 70 pc. \citet{corradi04} using Str\"omgren 
photometry and high-resolution spectroscopy suggested the existence of
two sheet-like structures, one at $d \leq 60$ pc and another around 120 - 150 pc
that is supposedly the interaction zone between the bubbles.

To determine the distance of the interaction zone between the LB and Loop I
we investigated the interstellar reddening in the region defined by
the Galactic coordinates $250\degr \leq l \leq 50\degr$ and 
$ -60\degr \leq b \leq 60\degr$.  The  Str\"omgren $uvby$H$\beta$ 
data  was taken from the  ``General Catalogue of Photometric Data" (GCPD),
compiled by \citet{hauck98}. The initial sample is composed of 8430 stars.

The determination of the intrinsic stellar parameters, colour excesses and 
distances are described in Sect. 2.  The data, error analysis and the application 
of the exclusion criteria are described in Sect. 3. Limiting magnitude effects on 
$E(b-y)$ and distances are described in Sect. 4, and colour excess diagrams are 
used in Sect. 5 to investigate the reddening distribution. The ring-like feature 
is investigated in  Sect. 6. A discussion of the results is given in Sect. 7 and the
conclusions are summarized in Sect. 8.


\section{Intrinsic stellar parameters, colour excess and distance}

The determination of the distances and  $E(b-y)$ colour excesses was  done using
the calibrations of \citet{crawford75} and \citet{olsen88} for the F-type stars, 
and \citet{crawford78,crawford79} for the  B-type and A-type stars, respectively.

The standard values of the physical parameters were obtained through interpolation
in the  standard relations of each spectral type.  The error determination of the 
intrinsic stellar parameters, colour excesses and distances have been obtained
for each  star individually following  the method suggested by \citet{knude78} 
complemented with the interpolation error, for  greater reliability.  

In a generic way for the colour excess we have:
$$E(b-y) = (b-y) - (b-y)_{0}$$ where the intrinsic colour $(b-y)_{0}$ is calculated 
according to the spectral type and $(b-y)$ is the measured colour index.
In the following subsections we present how the intrinsic stellar parameters,
colour excesses and distances are calculated for each spectral type. The error 
determination steps are also described.

\subsection{ F-type stars }

For the F-type stars intrinsic colour $(b-y)_{0}$ we used the calibration established by \citet{olsen88}. 
The procedure is used iteratively until convergence set by 
$\Delta (b-y)_{0} \leq 0.0001$ is reached. In the first iteration  the measured values are used.
The intrinsic colour  is calculated as:
\begin{eqnarray} (b-y)_{0} = 0.217 + 1.34\Delta\beta + 1.6(\Delta\beta)^{2} + C\delta c_{0} - (0.16 \nonumber \\
 \    +4.5\delta m_{0} +3.5\Delta\beta)\delta m_{0}; \hspace{1.0cm} for \ \delta m_{0} < 0.060.
\end{eqnarray}
\begin{eqnarray} (b-y)_{0} = 0.217 + 1.34\Delta\beta + 1.6(\Delta\beta)^{2} + C\delta c_{0}  \nonumber \\ 
 \  - (0.24\delta m_{0} + 0.035); \hspace{0.6cm}  for \ \delta m_{0} \geq 0.060,
\end{eqnarray}
where  $\Delta\beta = 2.72 - \beta$. The reddening free parameters are
 $\delta m_{0} = \delta m_{1} + 0.32E(b-y)$ and
 $\delta c_{0} = \delta c_{1} - 0.20E(b-y)$.  Stars with $\delta m_{0} > 0\fm135$,
which characterize the most extreme population II, were excluded from our analysis. 
The C factor defined by $$C = 4.9\Delta\beta + 32.2\delta m_{0} - 262.0(\delta m_{0})^{2} - 1.31$$
has to obey the following boundary conditions:

\begin{table}[htb]
\begin{center}
\begin{tabular}{llcr|l}
\hline
\multicolumn{4}{c|}{boundary condition} & C\\
\hline
if & $C > 1.6\Delta\beta$& and & $0.05 \leq \delta m_{0} \leq 0.09$&then C= $ 1.6\Delta\beta$ \\ [3pt]
if & $C \leq 0.013$& and &$ \delta m_{0} > 0.02$ &then C= 0.013 \\  [3pt]
if & $C \leq -0.05$& & &then C= -0.05 \\
\hline
\end{tabular}
\end{center}
\end{table}

The errors of the intrinsic stellar parameters are obtained as:
\begin{equation} \sigma_{(b-y)_{0}} = [(F_{\beta}\sigma_{\beta}^{obs})^{2} + (F_{\delta m_{0}}\sigma_{\delta m_{0}})^{2} + (F_{\delta c_{0}}\sigma_{\delta c_{0}})^{2}]^{1/2};
\end{equation}
where $$F_{\beta} = \frac{\partial (b-y)_{0}}{\partial \beta}, \hspace{0.5cm}
 F_{\delta m_{0}} = \frac{\partial (b-y)_{0}}{\partial \delta m_{0}},
  \hspace{0.5cm} F_{\delta c_{0}} = \frac{\partial (b-y)_{0}}{\partial \delta c_{0}},$$
$\sigma_{\beta}$ being  the measurement error in $\beta$ and
 $$\sigma_{\delta m_{0}} = [(\sigma_{m_{1}}^{obs})^{2} + (\frac{\partial m_{0}}{\partial \beta}\sigma_{\beta}^{obs})^{2} + (0.32\sigma_{E(b-y)})^{2}]^{1/2},$$
$$ \sigma_{\delta c_{0}} = [(\sigma_{c_{1}}^{obs})^{2} + (\frac{\partial c_{0}}{\partial \beta}\sigma_{\beta}^{obs})^{2} + (0.2\sigma_{E(b-y)})^{2}]^{1/2},$$
where $\sigma_{m_1}^{obs}$ and $\sigma_{c_1}^{obs}$ are the measurement errors in $m_1$ and $c_1$, respectively.
The error in the colour excess is calculated as:
\begin{equation} \sigma_{E(b-y)} = [(\sigma_{(b-y)}^{obs})^{2} + (\sigma_{(b-y)_{0}})^{2}]^{1/2},
\end{equation}
where $\sigma_{(b-y)}^{obs}$ is the measurement error in $(b-y)$ and $\sigma_{(b-y)_{0}}$ is calculated in
an interactive way until convergence in $\sigma_{E(b-y)}$.

For the calculation of the absolute magnitude we used the standard relation from \citet{crawford75}: 
\begin{equation} M_{V} = M_{V}^{std} - (9.0 + 20.0\Delta\beta)\delta c_{0}.
\end{equation}

Therefore,  the resulting error in the absolute magnitude is calculated as:
\begin{eqnarray}( \sigma_{M_{V}})^{2} = (\sigma_{M_{V}}^{std})^{2} + (20\delta c_{0}\sigma_{\beta}^{obs})^{2} + ((9 + 20\Delta_{\beta})\sigma_{\delta c_{0}})^{2} \nonumber \\
 + (\frac{\partial M_{Vzams}}{\partial \beta} \times\sigma_{\beta}^{obs})^{2};
\end{eqnarray}
where $\sigma_{M_{V}}^{std} = 0.25$  is the calibration error.
The interpolation error of $M_{V_{zams}}(\beta)$ is taken into account by the last term on the right side of Eq.(6).

\subsection{A-type stars }

For the A stars intrinsic colour $(b-y)_{0}$ we used the calibration established by \citet{crawford79}.
The calibration is  used iteratively until convergence in $(b-y)_{0}$. The intrinsic colour is  calculated as:

\begin{equation} (b-y)_{0} = 2.946 - \beta - 0.1\delta c_{0} - 0.25\delta m_{0}; \hspace{0.4cm} if \ \delta m_{0} < 0.0.
\end{equation}
\begin{equation} (b-y)_{0} = 2.946 - \beta - 0.1\delta c_{0}; \hspace{1.9cm} if \ \delta m_{0} > 0.0.
\end{equation}

For the   $(A3-A9)$ type stars, the indexes have the same meanings as for the  F-type  stars.
To the $A$ stars of the intermediate group ($A1-A2$), the Balmer jump that is a function of the surface gravity
also becomes affected by the temperature; thus the calibrations for this group
are not very accurate. Therefore, in this work  they have been excluded.   

The intrinsic colour  error  is calculated in
an interactive way until convergence in $\sigma_{E(b-y)}$. Since $\sigma_{\delta m_{0}}$ and $\sigma_{\delta c_{0}}$ 
depend on $\sigma_{E(b-y)}$, the equations are: 

\begin{equation} \sigma_{(b-y)_{0}} = [(\sigma_{\beta}^{obs})^{2} + (0.25\sigma_{\delta m_{0}})^{2} + (0.1\sigma_{\delta c_{0}})^{2}]^{1/2}; \hspace{0.2cm}  if \ \delta m_{0} \leq 0.0. \
\end{equation}
\begin{equation} \sigma_{(b-y)_{0}} = [(\sigma_{\beta}^{obs})^{2} + (0.25\sigma_{\delta m_{0}})^{2}]^{1/2}; \hspace{1.7cm} if \ \delta m_{0} > 0.0.
\end{equation}

The values of  $\sigma_{\delta m_{0}}$ and $\sigma_{\delta c_{0}}$ are calculated in the same way as for the F-type stars.

The absolute magnitude is calculated with the following equation:

\begin{equation} M_{V} = M_{V}^{std} - 9\delta c_{0} .
\end{equation}  

 The error in the absolute magnitude is calculated as:
\begin{equation} \sigma_{M_{V}} = [(\sigma_{M_{V}}^{std})^{2} + (9\sigma_{\delta c_{0}})^{2} + (\frac{\partial M_{Vzams}}{\partial \beta}\times\sigma_{\beta}^{obs})
^{2}]^{1/2},
\end{equation}
where  $\sigma_{M_{V}}^{std} = 0.30$  and the last term
is the interpolation error of  $M_{Vzams}(\beta)$.

\subsection{B-type stars}

In these stars the $\beta$ index is related to the luminosity and $c_{0}$ to the effective temperature.
The $(b-y)_{0}$, $\beta_{ZAMS}$ and $m_{0}$ values are interpolated from $c_{0}$ 
through the calibration established by \citet{crawford78}. The calibration is  used iteratively until
convergence in $c_{0}$.

The error in the intrinsic colour is calculated by:
\begin{equation} (\sigma_{(b-y)_{0}})^{2} = (\frac{\partial (b-y)_{0}}{\partial c_{0}}\times\sigma_{c_{0}})^{2};
\end{equation}
where $\frac{\partial (b-y)_{0}}{\partial c_{0}}$ is the derivative of the polynomial generated for
 $(b-y)_{0}$ as a function of $c_{0}$ from the standard relation.
For $\sigma_{c_{0}}$ we have:

\begin{equation}(\sigma_{c_{0}})^{2} = (\sigma_{c_{1}}^{obs})^{2} + (0.2\sigma_{E(b-y)})^{2};
\end{equation}
where $\sigma_{c_{1}}^{obs}$ is the  observed $c_{1}$  error and $\sigma_{E(b-y)}$ is calculated as described
in sect. 2.1 (Eqs 3 and 4).

The absolute magnitude is calculated by the following equations:
\begin{equation}
 M_{V} = M_{V}^{std} - 10(\beta_{ZAMS} - \beta) \hspace{1.5cm} for \ 0.20\leq \ c_{0} \leq \ 0.90
\end{equation}
\begin{equation}
M_{V} = M_{V}^{std} \hspace{2.5cm}  for \
c_{0} < 0.20.
\end{equation}
 Stars with $c_{0} > 0.90$ were excluded from our analysis.

The $M_{V}$ error is calculated as: 
\begin{equation} (\sigma_{M_{V}})^{2} = (\frac{\partial M_{V}}{\partial \beta}\times\sigma_{\beta}^{obs})^{2} + (\sigma_{M_{V}^{std}})^{2};
\end{equation}
where $\sigma_{M_{V}^{std}}=0.20$.

 The  $M_{V}(\beta)$ reference lines of  \citet{crawford78} are in good agreement with the trigonometric
luminosities, as confirmed in the test of the photometric distances of the B III, IV and V type stars via the Hipparcos
parallaxes, done by \citet{kaltcheva98}.

\subsection{Distances}

The  distance does not depend on the spectral type, being calculated as:
\begin{equation} log \ d = (V - M_{V} - 4.3E(b-y) + 5)/5,
\end{equation}
 assuming the standard extinction law (R$_{V}(uvby) = 4.3$.

The distance error is calculated by taking into account the dependences
 on $V$, $M_{V}$ and $E(b-y)$.
Thus we have:
\begin{equation} (\sigma_{d})^{2} =  (F_{V}\sigma_{V})^{2} + (F_{M_{V}}\sigma_{M_{V}})^{2} + (F_{E(b-y)}\sigma_{E(b-y)})^{2};
\end{equation}
where $F_{V} = (d/5)ln10$, \hspace{0.5cm}$F_{E(b-y)} = -4.3F_{V}$, \hspace{0.5cm}$F_{M_{V}} = -F_{V}$.

 It could be argued that the distances would be underestimated in those directions
where higher Rv values might be found, particularly  some directions towards the Galactic
center. However, as noted by \citet{fitz2007}, the regions where Rv is greater than 4.3
are very few, small and extremely localized.

To give an idea of how the distances would be affected by the difference in R$_{V}$ 
we calculated the percentual distance difference $\Delta$d, taken for
$E(b-y) = 0\fm100$, as shown in Table \ref{rv}.  The lowest R$_{V}$ values would take 
into account the higher X-ray and far-UV energy density inside the bubbles while the 
highest values would be associated with those very dense cores.  
Our choice for R$_{V}$ is shown to be appropriate when the photometric distances 
are compared to the Hipparcos distances in Sect. 3.2.

\begin{table}[htb]
\caption{ Effect of different R$_{V}$ values on the adopted distances,
taken  at a colour excess $E(b-y) = 0\fm100$. The plus and minus signs indicate whether the  
 calculated distance is larger  or smaller,  respectively. }
\label{rv} 
\begin{center}
\small{
\begin{tabular}{c|c|c|c|c|c|c|c}
\hline
R$_{V}$ (UBV)& 1.5 & 2.0 & 2.5 & 3.1  & 3.5 & 4.0 & 5.0\\
\hline
R$_{V}$ ($uvby$)& 2.1 & 2.8 & 3.5 & 4.3 & 4.9 & 5.6 & 6.9 \\
\hline
$\Delta$d & -10\% & -7\% & -4\% & - & +3\% & +6\% & +13\% \\
\hline
\end{tabular}
}
\end{center}
\end{table}

\section{The data}

In order to obtain the interstellar reddening towards the interaction zone 
 we used the $uvby$H$\beta$ data from the 
``General Catalogue of Photometric Data" (GCPD), compiled by \citet{hauck98}. We selected
the stars with a  complete set of data (V, $b-y$, $m_{1}$,  $c_{1}$, $\beta$),
eliminating the stars  classified as doubles, variables and peculiars in the GCPD. 

The individual errors of the measured values ($\sigma_{V}^{obs}$, $\sigma_{b-y}^{obs}$,
$\sigma_{m_{1}}^{obs}$,  $\sigma_{c_{1}}^{obs}$, $\sigma_{\beta}^{obs}$) were taken from 
the original papers. Whenever possible the average errors were used.

The initial sample has  8430 stars covering the region defined by the Galactic coordinates:
$250\degr\leq l \leq 50\degr$ and $ -60\degr \leq b \leq 60\degr$.

\subsection{Error analysis}

\begin{figure}[htb]
\begin{center}
\includegraphics[width=\linewidth]{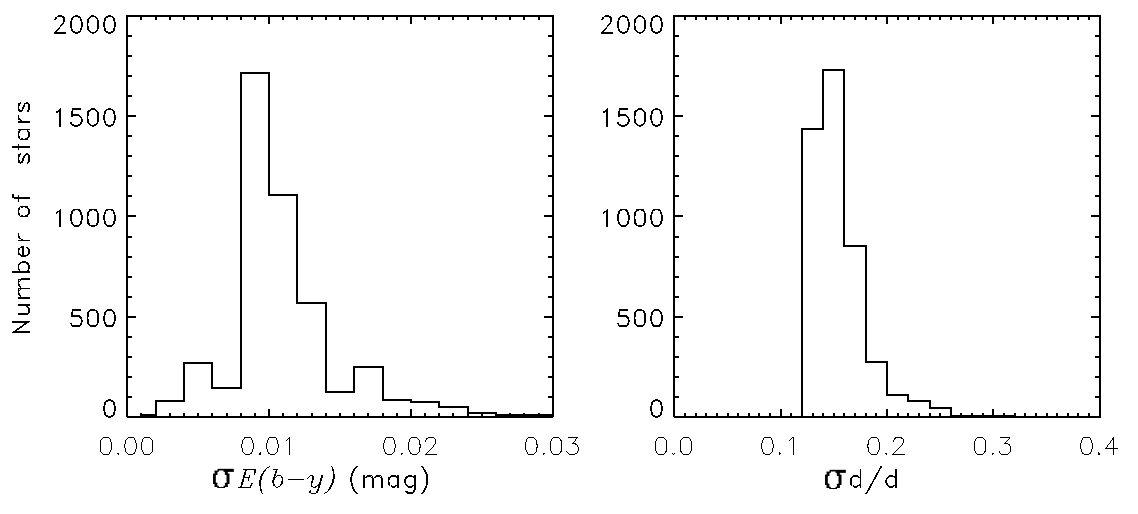}
\caption{Histograms for $\sigma_{E(b-y)}$ and $(\sigma_{d}/d)$. Notice that $\sigma_{E(b-y)} \approx 0\fm010$
and $\sigma_{d}/d \approx 0.15$. The accepted values were limited to $\sigma_{E(b-y)} \leq  0\fm025$
and $\sigma_{d}/d \leq 0.30$.  The cut-off in the histograms 
 is due to the distribution  of the measureament errors.}
\label{fracdist}
\end{center}
\end{figure}

As can be seen on the left side of Fig. \ref{fracdist} the histogram  of the error in $E(b-y)$,
namely $\sigma_{E(b-y)}$, shows  greater concentration around $0\fm010$, with  a dispersion of  
$\approx 0\fm007$.  To assure the reliability of our results we decided to accept only  
$\sigma_{E(b-y)} \leq 0\fm025$, that is limited to the average of $\sigma_{E(b-y)}$ plus 2 times 
the dispersion.

Since the  error of the distance ($\sigma_{d}$) is proportional to the distance of the star, 
we have used the relative error $\sigma_{d}/d$  in our analysis. As can be seen in the 
histogram of $\sigma_{d}/d$ on the right side of Fig. \ref{fracdist}, the  $\sigma_{d}/d$ 
shows  higher concentration around  0.15, with a dispersion of $ \approx 0.075$. Consistently 
with $\sigma_{E(b-y)}$, the accepted values were limited to $\sigma_{d}/d \leq 0.30$.

\subsection{Application of the exclusion criteria}

For a colour excess to be useful for interstellar medium studies  a set of selection criteria must be met. The  
exclusion photometric criteria are those proposed by \citet{olsen79}, \citet{crawford75,crawford78,crawford79}
and \citet{tobin}.

  For the B stars it should be kept in mind that possible Balmer line emission makes the absolute
magnitudes and thus the distances less precise. We have chosen the Tobin (1985) exclusion criteria,
even at a cost of excluding more than 40\% of the B-type stars of the available sample.

 We also used the  SIMBAD database at  CDS   
to verify the characteristics of the stars in our sample and 
to eliminate any object classified  as double,
peculiar or variable.

\begin{figure}[htb]
\begin{center}
\includegraphics[width=0.8\linewidth]{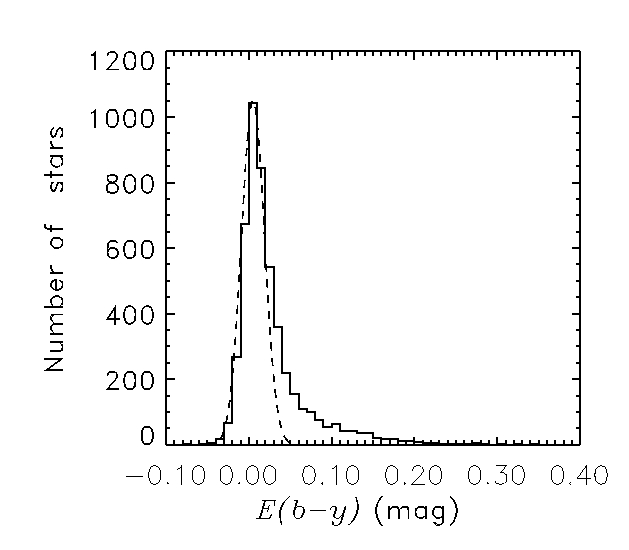}
\end{center}
\caption
{Histogram of the colour excess for the stars of our sample.
 The Gaussian curve indicates that there is  no  zero point effect.}
\label{ebyhisto5}
\end{figure}

 In  each  set of  $E(b-y)$ the unreddened values are expected to be clustered around zero.
As can be seen in the histogram of $E(b-y)$ shown in  Fig. \ref{ebyhisto5},  the Gaussian
curve indicates that there is no zero point effect.
The center of the Gaussian is located at $0\fm005$ and $\sigma = 0\fm020$ and 
as can be seen in Fig. \ref{ebyhisto5} the number of stars below n*sigma are within 
expectation. 

Very few higher negative colour excess values  (i.e. $E(b-y) \leq -0\fm011$) are still present in the final sample
after the application of the exclusion criteria. Most of them are within 1.3 $\times$  the average of $\sigma_{E(b-y)}$.
Although they indicate a possible star peculiarity we have kept them in the final sample but  
with a separate code indicated in the figures.

To  verify the precision of the photometric distances ($d_{Phot}$)
and to ensure the validity of R$_{V} = 4.3$, we compared our  results  with the
trigonometric  distances ($d_{Hip}$)  determined by the Hipparcos satellite.
We considered only  stars with $\sigma_{\pi} /\pi \leq 0.30$, that is, the same 
error range  used for the photometric distances. We eliminated those stars whose 
$d_{Hip}$ did not coincide with $d_{Phot}$ within the error range. The  stars 
eliminated by this criterion did not show any trend with spatial location.

\begin{figure}[htb]
\begin{center}
\includegraphics[width=\linewidth]{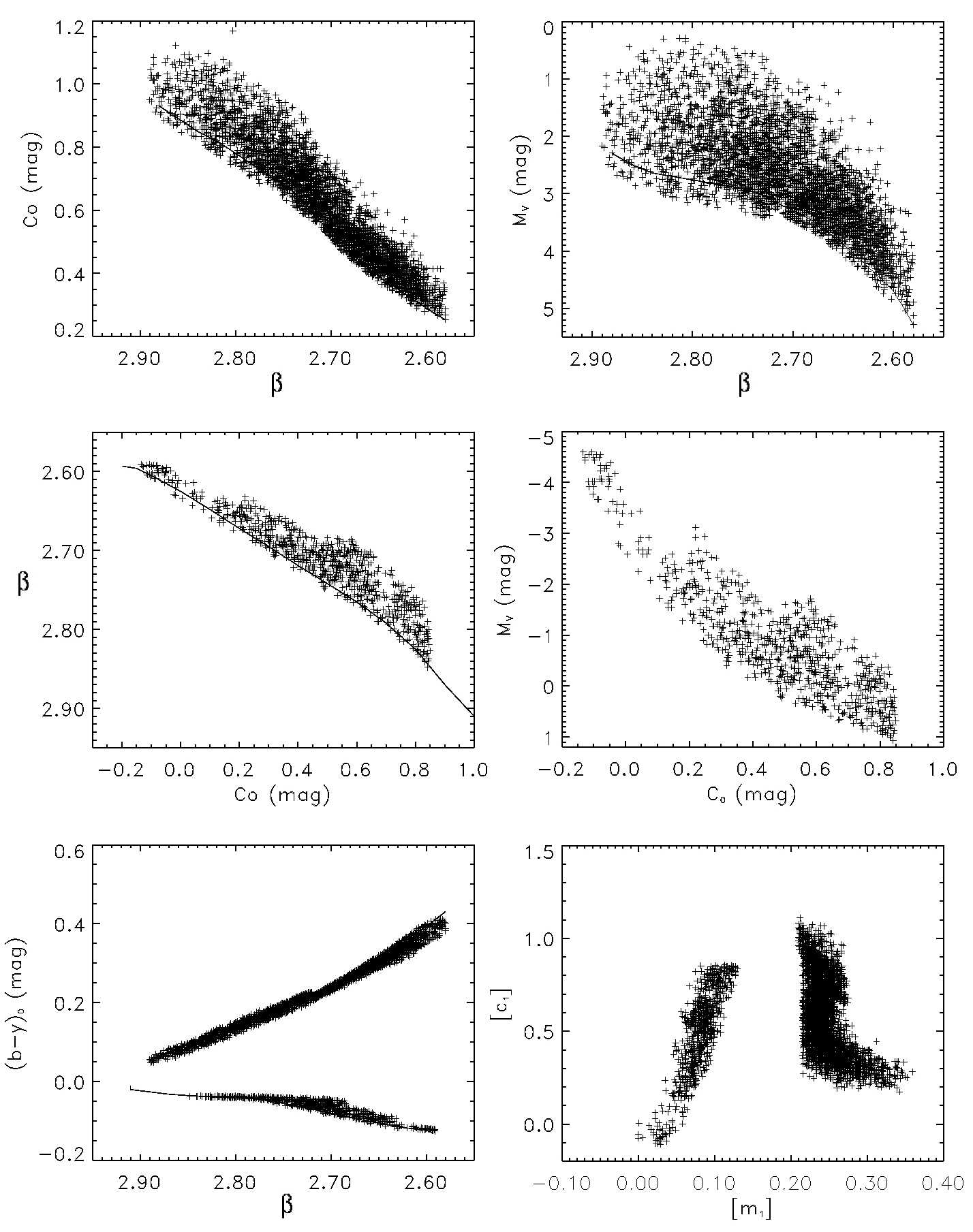}
\end{center}
\caption{ ({\it Top-left})  $c_{0} \ vs. \ \beta$ and ({\it Top-right}) M$_{V} \ vs. \ \beta$ 
diagrams   for the  AF stars.  ({\it middle-left}) $\beta \ vs. \ c_{0}$,
({\it middle-right}) M$_{V} \ vs. \ c_{0}$ diagrams for OB select stars. ({\it bottom-left}) 
$(b-y)_{0} \ vs. \ \beta$ and ({\it bottom-right}) $[c_{1}] \ vs. \ [m_{1}]$ diagrams for 
the final sample. Notice that we do not observe any  major systematic  disagreement with 
the standard relation.  Further details in the text. 
}
\label{cob}
\end{figure}

In  Fig. \ref{cob} we show the physical stellar parameters
of the selected stars with the superposition of the standard relation curves.
We do not observe any  major systematic  disagreement with the standard relation.
We also give the $[c_{1}]$ $vs.$  $[m_{1}]$ diagram of the selected stars.
The  stars of the intermediate group (A1-A2) were eliminated, as were 
stars brigther than  luminosity class III and  spectral types G, K and M.

\begin{figure}[htb]
\begin{center}
\includegraphics[width=0.9\linewidth]{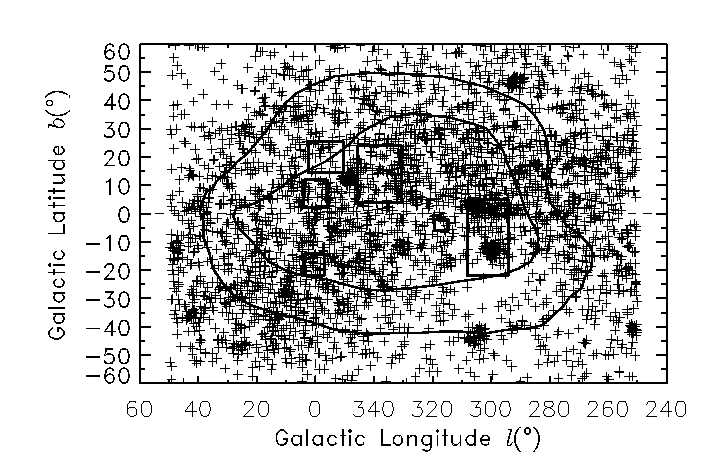}
\end{center}
\caption
{ Distribution of the final sample over  the studied area. The ring-like feature 
proposed by \citet{egger95} is also plotted for reference. As can be seen the 
part of the ring located at $290\degr \leq l \leq 310\degr$ and
$-35\degr \leq b \leq -25\degr$ is less densely populated.}  
\label{geral}
\end{figure}

Our final sample has 4346 stars up to 500 pc from  the Sun.
The distribution of the final sample over the studied area is given in Fig. \ref{geral}
and the ring-like feature proposed by \citet{egger95} is also plotted  for reference.  
As can be seen the part of the ring located at $290\degr \leq l \leq 310\degr$ and
$-35\degr \leq b \leq -25\degr$ is less densely populated.

\section{Limiting magnitude effects on $E(b-y)$ and distances}

In a magnitude  limited sample, the reddening data are expected to be complete to
a maximum observable colour excess given at each distance \citep{knude87}.

\begin{figure}[htb]
\begin{center}
\includegraphics[width=\linewidth]{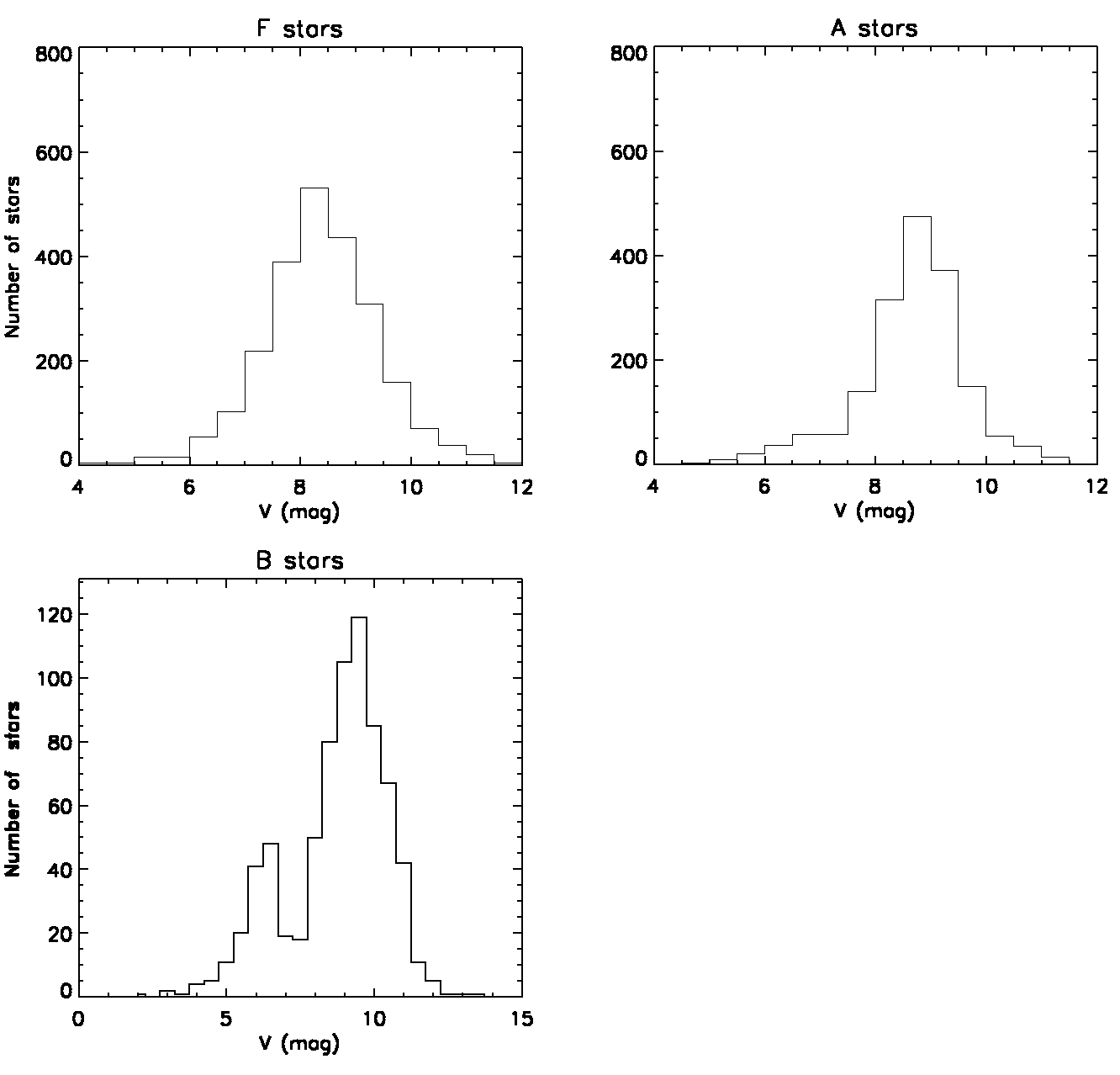}
\end{center}
\caption{Histograms of $V$  magnitude   
 for the  F-  ({\it top-left}), A- ({\it top-right}) and 
  B-type  stars  ({\it bottom}). Note the decrease in the number of stars fainter than
$V\approx 9\fm0$ for the F-, $V\approx 9\fm5$ for the A-  and $V\approx 10\fm0$ for the 
B-type stars, meaning that the final sample may be complete, in magnitude,
only down to these values.}
\label{histomag}
\end{figure}

 The histograms of the $V$ magnitudes, shown in  Fig. \ref{histomag}, indicate
that there is a decrease in the number of stars fainter than  $V\approx 9\fm0$ 
for the F-, $V\approx 9\fm5$ for the A-  and $V\approx 10\fm0$ for the B-type stars.

As our sample may be complete only down to these values, it would be important 
to see if this limit has any effect on the  colour excess distribution.

\begin{figure}[htb]
\begin{center}
\includegraphics[width=0.8\linewidth]{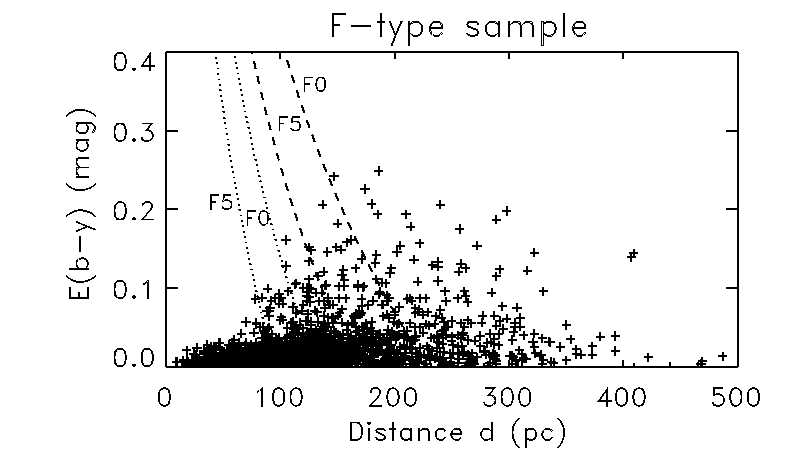}
\includegraphics[width=0.8\linewidth]{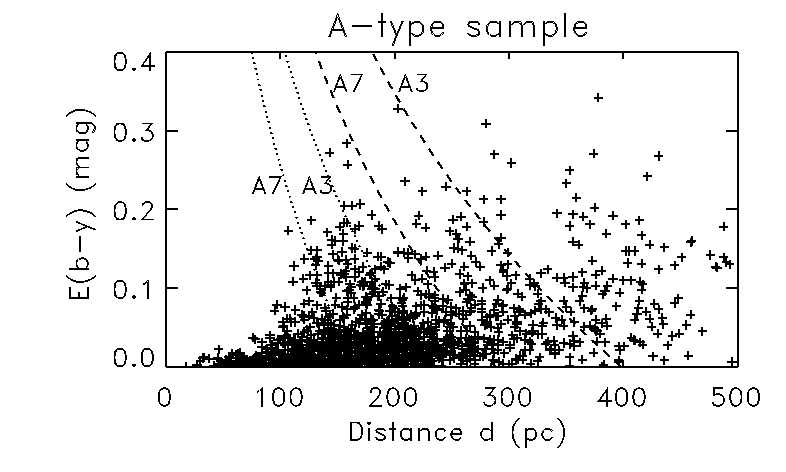}
\includegraphics[width=0.8\linewidth]{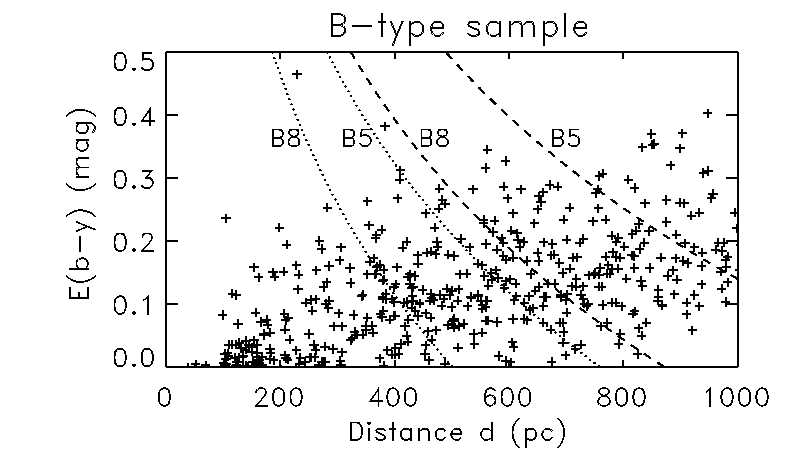}
\end{center}
\caption{ $E(b-y)$  $vs.$  d (pc)  diagram  for the F-, A- and B-type stars of the final sample
(from  top to  bottom). The dotted and dashed curves indicate the maximum detectable
$E(b-y)$ as a function of the distance for classes  F0\,V, F5\,V, A7\,V, A3\,V, B8\,V and B5\,V stars
with magnitude V =  $8\fm3$ and  V =  $9\fm5$. 
Assuming they exist, the greater reddenings could be picked up by the intrinsically brighter stars of the sample.}
\label{edcurva}
\end{figure}

 The $E(b-y)$ $vs.$ distance diagrams for the F-, A- and B-type stars of the final sample are shown in 
Fig.  \ref{edcurva}. They illustrate that higher values of colour excesses could have been detected,
if they exist. In these diagrams the dotted and dashed curves indicate the maximum detectable
$E(b-y)$ as a function of the distance for classes  F0\,V, F5\,V, A7\,V, A3\,V, B8\,V and B5\,V stars
with magnitude V =  $8\fm3$ and  V =  $9\fm5$, respectively.

 When the F-type stars are no longer capable of picking up the larger reddenings, the A- and
B-type stars, being intrinsically brighter, exist in  enough numbers to detect such larger reddenings,
if present.

According to \citet{egger95}, the column  density of neutral hydrogen in the line of sight of the ring
jumps from $10^{20}$ cm$^{-2}$ to $7 \times 10^{20}$ cm$^{-2}$ at $d \approx 70$ pc, corresponding to
$E(b-y)$ from  $\approx 0\fm015$ to $\approx 0\fm100$, if it is assumed that the standard relation
between $E(b-y)$ and $N_{H}$ \citep{knude78b} is valid.

 Since the data sample is drawn from a general photometric compilation, and the original data
was obtained for different purposes we have also to verify whether the data are pertinent to our analysis.
In particular, we need to know if the final sample is adequate for measuring the
possible existence of an extinction jump from $E(b-y)\approx0\fm015$ to $E(b-y)\approx 0\fm100$ around 100 pc,
which would correspond to the bubbles' ring.
We refer the reader to Sect. 6.4 where we show separate histograms of the V magnitudes of the
$\beta$ index and the absolute magnitude M$_{V}$ for each area used to investigate the existence of
the bubbles' interface.

\section{The interstellar reddening towards the interaction zone}

To investigate the interstellar reddening distribution towards the interaction zone of the two bubbles we will plot
the colour excess as a function of the Galactic coordinates ($l$, $b$). The colour excesses are divided according to
the symbols and colors shown in Table \ref{lb}. 

\begin{table}[htb]
\begin{center}
\caption{Symbols and colors used in the reddening  analysis.}
\label{lb}
\begin{tabular}{l|c|c|c}
\hline
$E(b-y)$ (mag)& Symbol&Color&B \& W\\
\hline
$\leq$ -0.011&open triangle&black&black \\
-0.011 - 0.010& square &yellow&light gray \\
0.010 - 0.020 &open square &green&light gray\\
0.020 - 0.040 &$\times$ &light blue&light gray \\
0.040 - 0.070 &$\times$ &purple&dark gray\\
0.070 - 0.100 & + &red&black\\
0.100 - 0.200 &circle&black&black\\
$\geq$ 0.200 &square&black&black\\
\hline
\end{tabular}
\end{center}
\end{table}

This division is based on the fact that typically stars with $E(b-y) \leq  0\fm010$ can be considered unreddened,
consistently with $\sigma_{E(b-y)} \approx 0\fm011$. Stars that present $E(b-y) \geq 0\fm040$ are probably screened
by at least one small diffuse interstellar cloud, whose mean colour excess is around $0\fm030$ according
 to \citet{knude78}. Stars showing $E(b-y) \geq 0\fm100$ are certainly screened by at least one dense 
interstellar cloud.  

\begin{figure*}[htp]
\includegraphics[width=\linewidth]{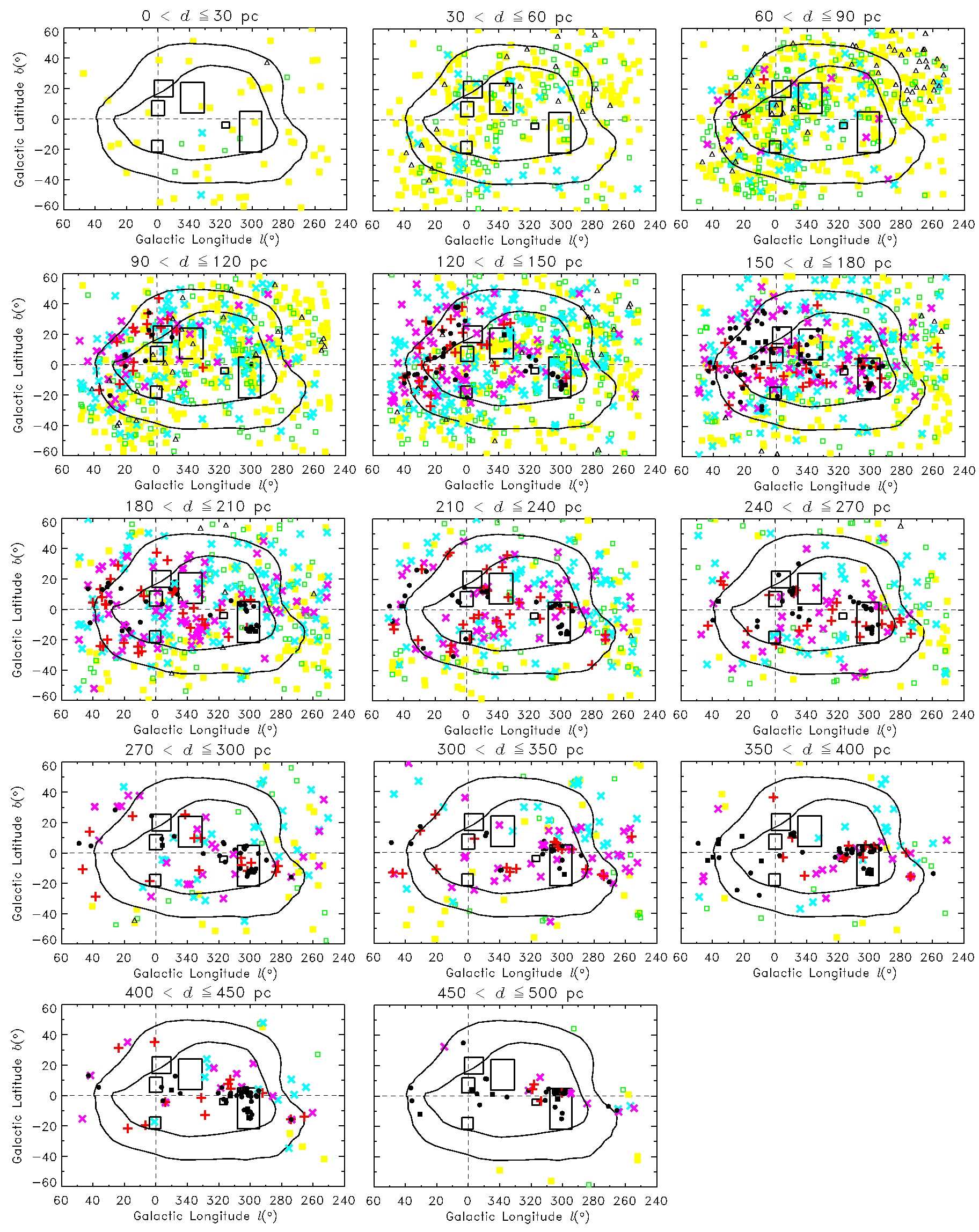}
\caption{Diagrams of colour excess distribution divided by intervals of the  distance to $d = 500$ pc.
Our data show that up to 60 pc the $E(b-y)$ values are below $0\fm040$ in all directions,
$0\fm020$ being a typical value.
The first stars with reddening corresponding to the ring column density -- $ E(b-y) \approx 0\fm100$ --
appear on the western side between $90 < d  \leq 120$ pc, whereas on the eastern side they appear
only after d$\approx$270 pc.}
\label{060}
\end{figure*}

\begin{figure*}[htb]
\includegraphics[width=\linewidth]{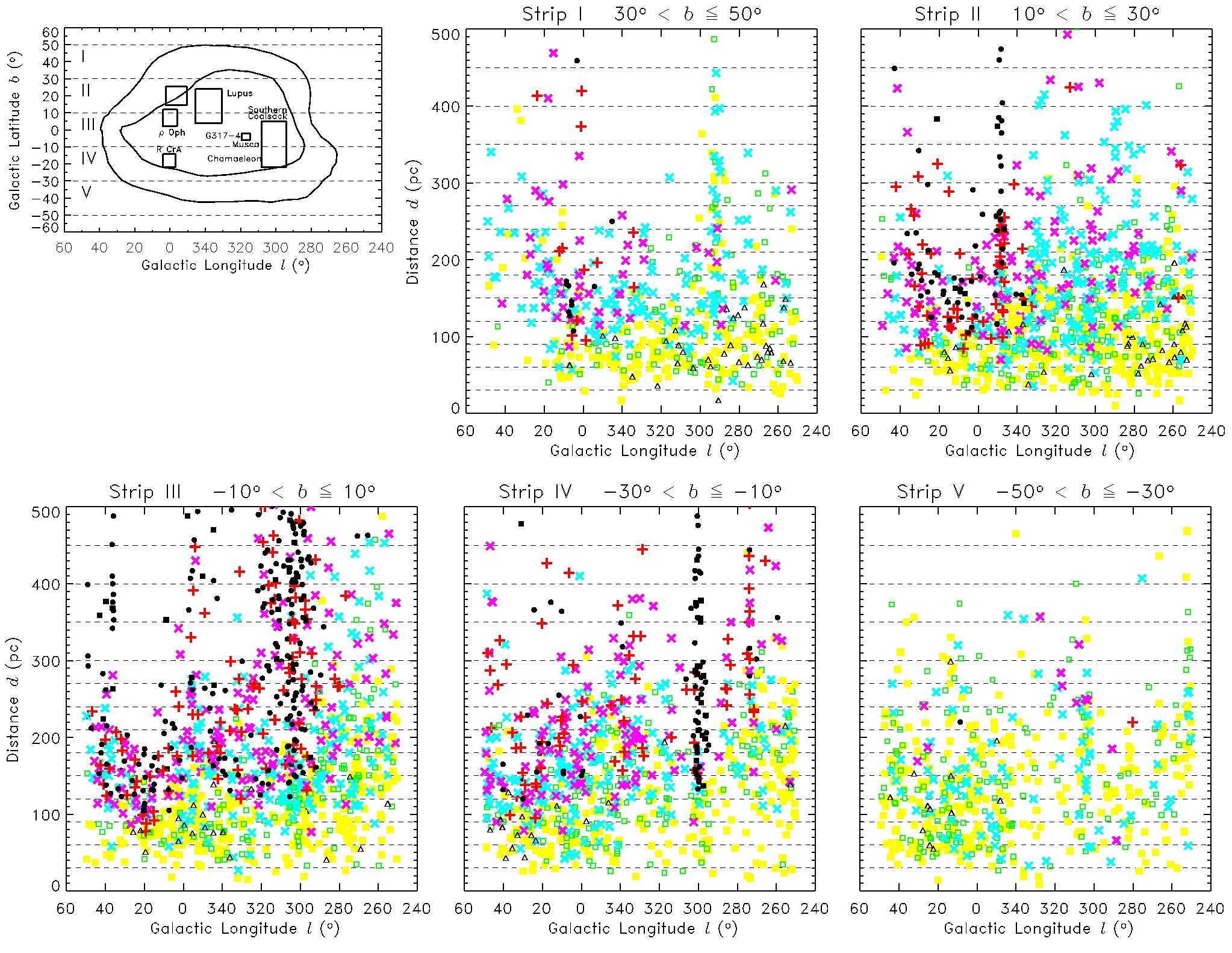}
\caption{Diagrams of the  distribution of  colour excess in Galactic latitude intervals.
Notice that the expected transition to $E(b-y) \approx  0\fm070 - 0\fm100$ occurs on the western
part at  90 - 120 pc, whereas on the eastern side it is not clearly seen before 280 pc.
Near the Galactic plane the presence of the dark clouds is clearly identified by $E(b-y) \geq 0\fm100$.
In the southernmost part of the ring the reddening remains very low, typically
 $E(b-y) = 0\fm020$ in all its extension, except towards the Mensa constellation
where a possible transition is observed at $d \approx 200$ pc.}
\label{dlong1}
\end{figure*}

Figure \ref{060} shows the stars according to their position in the sky and divided to intervals of 30 pc
to 300 pc, and by 50 pc intervals  to 500 pc. The ring-like feature proposed by \citet{egger95} and a
schematic contour of the  $\rho$ Oph, Lupus, R CrA, \mbox{G317-4}, Southern Coalsack, 
Chamaeleon and Musca dark clouds are also plotted in this figure for reference. For clarity 
the cloud's names are indicated in Fig. \ref{dlong1}.

 Our data show that up to 60 pc the $E(b-y)$ values are below $0\fm040$ in all directions, 
$0\fm020$ being a typical value.  The first stars with reddening corresponding to the ring 
column density -- $ E(b-y) \approx 0\fm100$ -- appear on the western side between 
$90 < d  \leq 120$ pc, whereas on the eastern side they appear only after d$\approx$270 pc.

On the western side most of stars beyond 120 pc show a colour excess between $0\fm070  < E(b-y) < 0\fm100$, 
whereas on the eastern side we still have many stars with $E(b-y) < 0\fm020$ up to 270 pc, where 
a definite transition occurs to  $E(b-y) \approx 0\fm070-0\fm100$.

From 100--180 pc we clearly see the presence of the dark clouds, as indicated by the black circles 
$E(b-y) > 0\fm100$ in Fig. 7. A thorough discussion will be given in the next section.

\subsection{Analysis of the reddening as function of the distance and the Galactic latitude}

Complementing the previous diagrams we will plot diagrams of d (pc) $vs.$
Galactic longitude $l(\degr)$ for five Galactic latitude strips of $20\degr$ width each.
These diagrams are presented in Fig. \ref{dlong1} together with the position  of each  
strip and are analysed in the next paragraphs.

\begin{itemize}
\item Region I ($30\degr < b \leq 50\degr$):

\smallskip

The majority of the stars in this direction show $E(b-y) < 0\fm020$ (yellow and green squares) up to 100 pc
on the western side ($ 340\degr < l < 50\degr$) and up to 150 pc on the eastern side ($ 340\degr > l > 250\degr$).
A transition to $E(b-y) \approx 0\fm040$ (purple crosses) occurs approximately at 120 pc only on the western side.
In the direction of the ring, the expected transition to $E(b-y) \approx  0\fm070 - 0\fm100$ (red plus signs) is not observed, 
except for a small region around $l \approx 10\degr$ associated with the $\rho$ Oph molecular cloud where $E(b-y) > 0\fm100$ 
(black dots).

\medskip

\item Region II ($10\degr < b \leq 30\degr$):

\smallskip
 A transition to $E(b-y) \approx  0\fm070 - 0\fm100$ (red plus signs) appears at about 90 pc between 
$350\degr \leq l \leq 30\degr$. However, in the eastern side a similar transition is not
observed. Even a minimum color excess cannot be clearly assigned, given the large number
of stars with $ E(b-y) \approx 0\fm020$ (yellow and green squares) up to at least  150 pc.
Still on the western side, several stars present $ E(b-y)> 0\fm100$ (black dots) from 100 - 150 pc indicating that
some parts of the Lupus and  Oph-Sgr molecular clouds have been crossed.
\medskip

\item Region III ($-10\degr < b \leq 10\degr$):

\smallskip
 A transition to $E(b-y) \approx  0\fm070 - 0\fm100$ (red plus signs) occurs at about 90 pc 
on the western side of the ring ($25\degr \leq l \leq 40\degr$), changing to 120 -- 150 pc 
in its central parts ($285 \degr \leq l \leq 25\degr$). A similar transition on its eastern 
side is not clearly observed up to $d \approx 280$  pc (no red plus signs are seen).
Between $250\degr \leq  l \leq 285\degr$, most stars show very low reddening $E(b-y) < 0\fm020$  
(yellow and green squares) up to $d \approx 240$ pc, with a hint of a minimum value of 
$E(b-y) = 0\fm040$ (purple crosses) from 180 pc. The directions with $E(b-y) > 0\fm100$ (black dots) 
indicate the presence of the molecular clouds; in the western side they are seen from 100 - 150 pc 
(Scutum and Lupus) whereas in the eastern side they are seen from 120 - 180 pc (G317-4, Coalsack, Musca). 

\medskip

\item Region IV ($-30\degr < b \leq -10\degr$):

\smallskip
 The reddening in this region follows the same trend as the previous strip. 
Remarkably, between $ 265\degr \leq  l \leq 290\degr$, the values of $E(b-y) \geq 0\fm100$ (black dots) only appear at
$d \approx 280$ pc. But, in the area internal to the ring we have the presence of the molecular clouds Sag-South, Aql-South
and R CrA at 120 -- 150 pc on the western side and Chamaeleon at $d \approx 140$ pc on the eastern side.

\medskip

\item Region V  ($-50\degr < b < -30\degr$):

\smallskip

 There is no sign of the transition to $E(b-y) \approx  0\fm070 - 0\fm100$ (no red plus signs). A hint of a transition to
$E(b-y) > 0\fm040$ (purple crosses) occurs at 200$\pm20$ pc, but is restricted to $270\degr \leq l \leq 310\degr$ and
$-45\degr \leq b \leq -33\degr$. This is about the same position and distance as the bright infrared filament observed by
\citet{penprase98}. The reddening remains very low in the whole area (yellow and green squares and light blue crosses), 
$E(b-y) = 0\fm020$ being a typical value.

\end{itemize}

 Thus, the expected transition to $E(b-y) \approx  0\fm070 - 0\fm100$ occurs in the western  part of the ring
at about 90 and 120 pc,  whereas on the eastern side it is not clearly seen before 280 pc.
Near the Galactic plane the presence of the dark clouds is clearly identified by $E(b-y) \geq 0\fm100$ on the western side at
about 100 -- 150 pc and a little further on the eastern side, at about 120 -- 180 pc. In the southernmost part of the ring
the reddening remains very low, typically  $E(b-y) = 0\fm020$ in all its extension, except towards the Mensa constellation
where a possible transition is observed at $d \approx 200$ pc. 

\section{On the existence of a ring-like feature}

\begin{figure*}[ht]
\centering
\includegraphics[angle=0,width=1.00\linewidth]{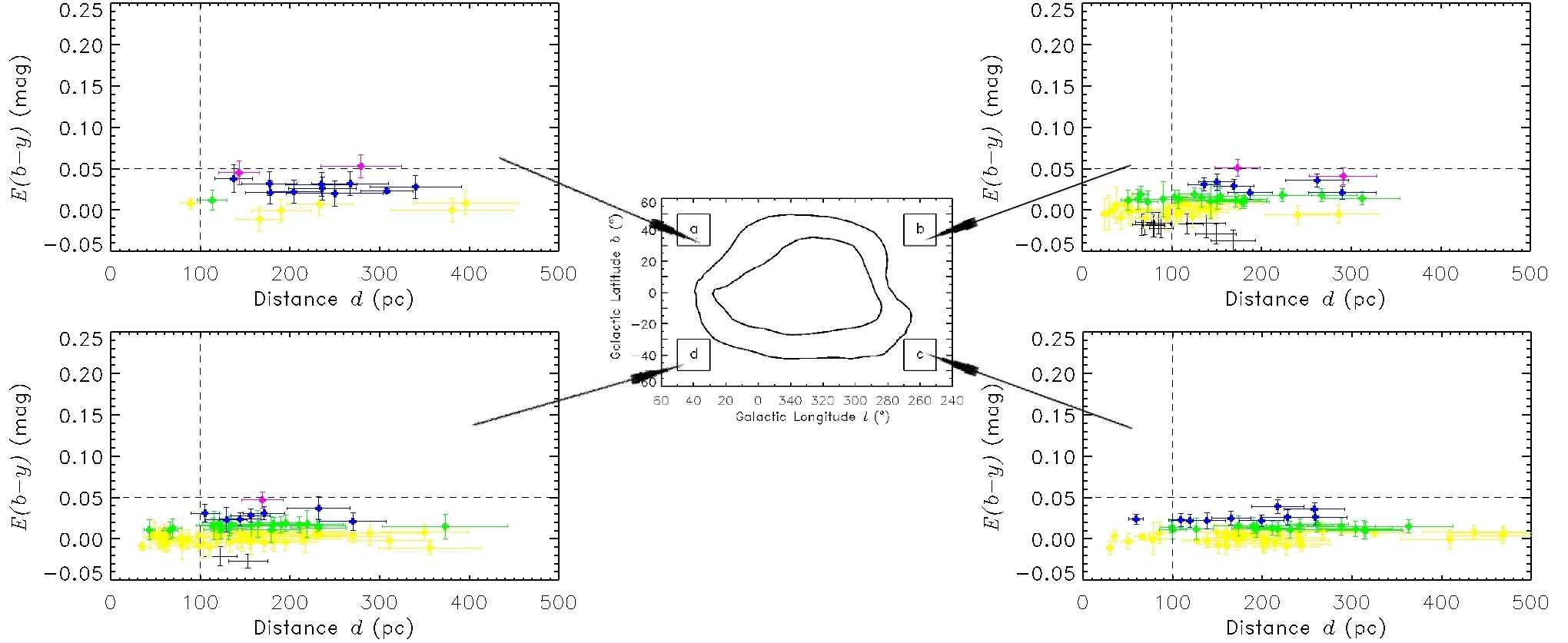}
\caption{$E(b-y)$ (mag) $vs.$ distance $d$ (pc) diagrams (areas outside the ring).
In the  areas outside the ring the colour excess remains low
($E(b-y) < 0\fm040$) up to the maximum distance of the stars of our sample, $0\fm020$ being a typical value.}
\label{edistext}
\end{figure*}

 The ring-like feature proposed by  \citet{egger95} would be the  most prominent characteristic
of the interaction zone.  To identify  the ring 
we  plot  diagrams of $E(b-y)$ (mag) $vs.$ $d$ (pc), with the respective errors,
for areas outside the ring, along the ring feature contour and inside the ring.
In these diagrams the horizontal dashed line  indicates 
  $E(b-y) = 0\fm050$ and the vertical dashed line $d = 100$ pc.

\subsection{ Analysis of the  $E(b-y)$ $vs.$ distance diagrams for  areas outside  the ring feature}

In Fig. \ref{edistext} (from ``a" to ``d") we present the 
$E(b-y)$ (mag) $vs.$ $d$ (pc) diagrams for  areas outside  the ring. The respective
coordinates are given in Table \ref{coordareas1}.

\begin{table}[htb]
\caption{Coordinates of the areas outside  the ring.}
\label{coordareas1}
\begin{center}
\begin{tabular}{l|c|c|c|c}
\hline
area&$l_{min}(\degr)$&$l_{max}(\degr)$&$b_{min}(\degr)$&$b_{max}(\degr)$\\
\hline
a&30 & 50 &30 & 50 \\
b&250 & 270 &30 & 50 \\
c&250 & 270 &-50 & -30 \\
d&30 & 50 &-50 & -30 \\
\hline
\end{tabular}
\end{center}
\end{table}

In the  areas outside the ring the colour excess remains low
($E(b-y) < 0\fm040$) up to the maximum distance of the stars 
of our sample, $0\fm020$ being a typical value.

\subsection{Analysis of the   $E(b-y)$  $vs.$ distance diagrams for areas along the ring}

\begin{figure*}[htb]
\begin{center}
\includegraphics[angle=-90,width=0.62\linewidth]{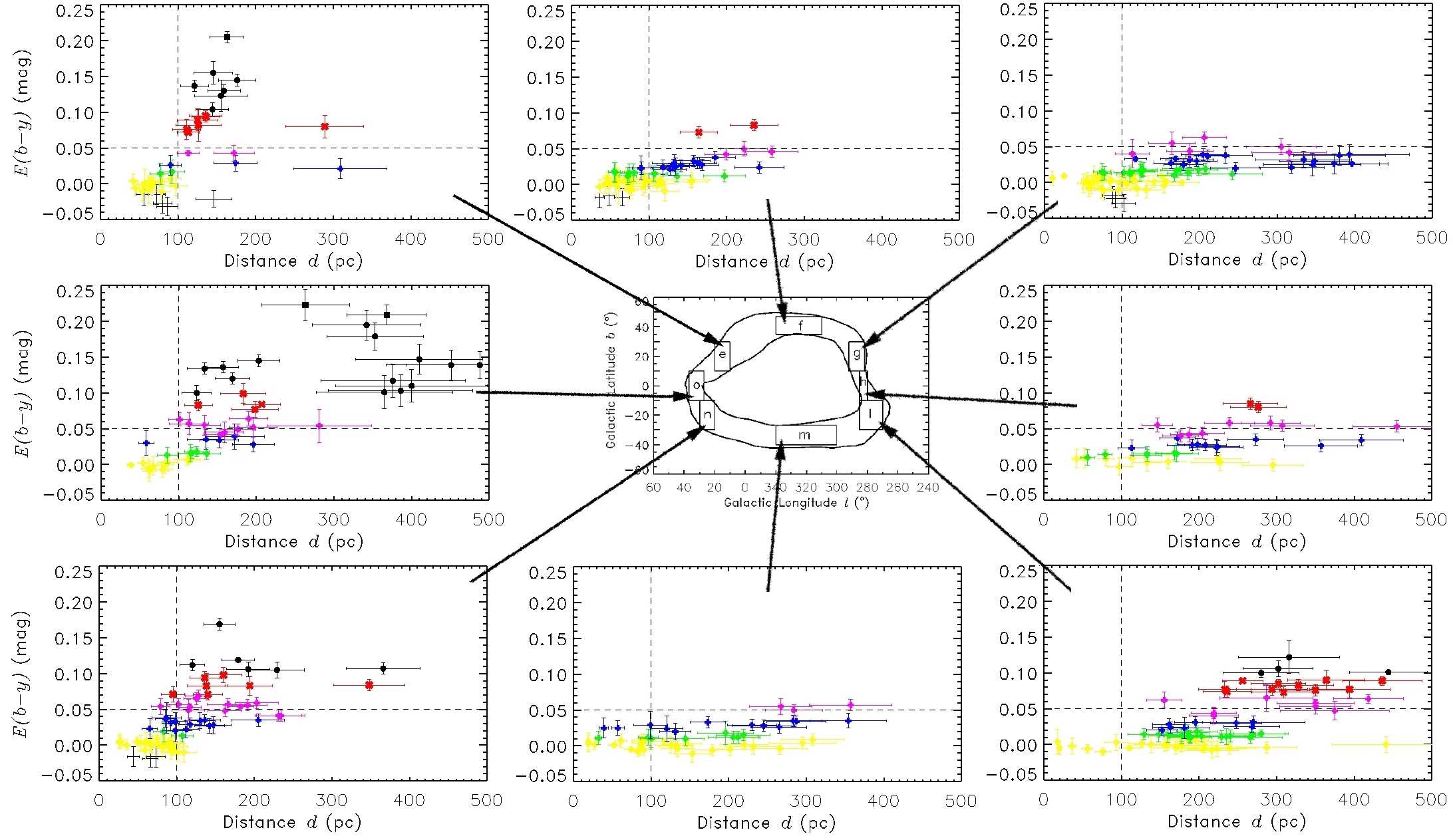}
\caption
{$E(b-y)$ (mag.)  $vs.$ distance $d$ (pc) diagrams (areas located along  the ring). On the
 western side of the ring the colour excess increase to values greater than $E(b-y) \geq 0\fm070$ 
at $d = 110\pm$20 pc whereas on the eastern side the same values of colour excess occur only in area ``l",
but at $d = 280\pm$50 pc. In the northern parts of the ring the colour excess change in a gradual way
(area ``f") while in its southern parts (area ``m") the colour excess remains low ($E(b-y) \leq 0\fm040$)
up to $d \approx 300$ pc. }    
\label{edistno}
\end{center}
\end{figure*}

In Fig. \ref{edistno} (from ``e" to ``o") we present the  
same diagrams for the areas located along the ring. The respective 
coordinates are given in Table \ref{coordareas2}.

\begin{table}[htb]
\caption{Coordinates of the  areas located along the ring.}
\label{coordareas2}
\begin{center}
\begin{tabular}{l|c|c|c|c}
\hline
area&$l_{min}(\degr)$&$l_{max}(\degr)$&$b_{min}(\degr)$&$b_{max}(\degr)$\\
\hline
e&10 & 20 &10 & 30 \\
f&310 & 340 &35 & 47 \\
g&282 & 292 &10 & 30 \\
h&280 & 285 &-10 & 10 \\
l&270 & 285 &-30 & -10 \\
m&300 & 340 &-40 & -27 \\
n&20 & 30 &-30 & -10 \\
o&27 & 37 &-10 & 10 \\
\hline
\end{tabular}
\end{center}
\end{table}

On the western side of the ring, in areas  ``e", ``n" and ``o", the first very reddened stars 
($0\fm070 \leq E(b-y) \leq 0\fm100$) appear at $ 110 \pm 20$ pc. 

Area ``e" contains the molecular cloud Oph-Sgr ($8\degr < l < 40\degr, 9 \degr < b < 24\degr$)
 and area ``n" contains the Sag-South and Aql-South ($27\degr < l < 40\degr, -21 \degr < b < -10\degr$).
In both areas there is a transition to $E(b-y) \geq 0\fm100$ at $130 \pm 20$ pc, indicating the distance
to these two objects. 

Near the Galactic plane, in the general direction of the Scutum dark cloud ($l = 25, b = 1$), area
``o" shows a first jump to higher colour excess values at $130 \pm 20$ pc and a second jump at $300 \pm 40$ pc.

In the  northernmost part of the ring, area ``f", the colour excess increases with the distance in a gradual way,
such that $0\fm020 \leq E(b-y) \leq 0\fm040$ becomes predominant only after $d = 120\pm$15  pc. 
In the eastern side of the ring, areas ``g" and ``h", the colour excess remains below $E(b-y) \leq 0\fm040$  
up to the maximum observed distance. Only very few stars show a reddening slightly greater than this
after $d = 180\pm$20 pc.

In  area ``l" the  reddened stars ($0\fm070 \leq E(b-y) \leq 0\fm100$) appear only at $d = 280\pm$50 pc.
 After this distance the reddening remains constant.
 Remarkably, this area contains the abovementioned infrared bright
filament towards the Mensa constellation that is also located at 230 $\pm$ 30 pc 
and has a {\rm H\mbox{\sc{I}}} column density
$N_{\rm HI} = \ (8.22 \pm 2.5) \times 10^{20}$ cm$^{-2}$ \citep{penprase98} and corresponding 
$E(b-y) = 0\fm120 \pm 0\fm040$.

The colour excess in the southern part of the ring, area ``m", remains low ($E(b-y) \leq 0\fm040$), at 
least up to $d \approx 300$ pc.
  
On the western side of the ring the distribution 
of the colour excess occurs in a very different way than
the eastern side (both to northern and southern areas).
Such results have led us to think that the annular ring feature may not be a real entity.

\begin{figure*}[ht]
\flushleft
\includegraphics[angle=0,width=1.00\linewidth]{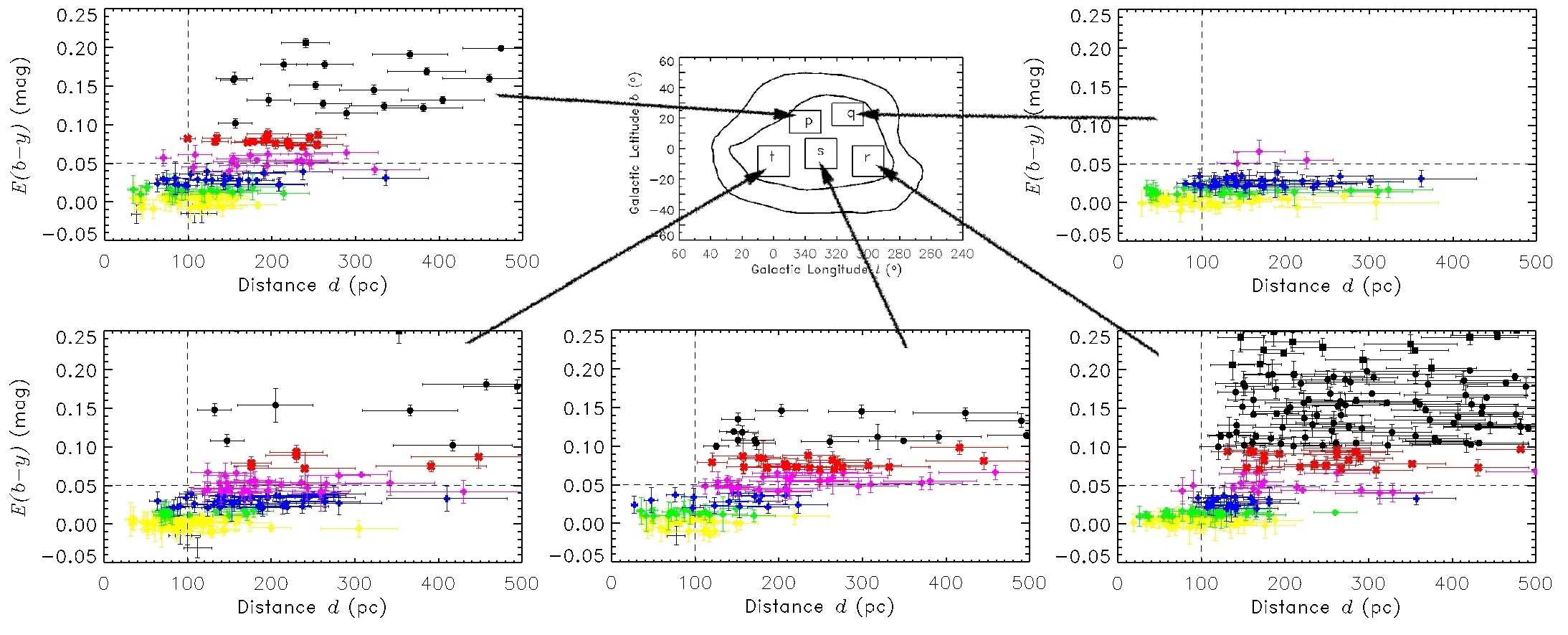}
\caption
{ $E(b-y)$ (mag.) $vs.$ distance  $d$ (pc) diagrams (areas inside  the ring).
The presence of the dark clouds, clearly identified by $E(b-y) \geq 0\fm100$,
can be seen on the western side at about 100 -- 150 pc and on the eastern side at about 120 -- 180 pc
in all areas, except for area ``q". Beyond these distance ranges the number
of unreddened stars decreases considerably indicating the location of these large dark cloud complexes.
}
\label{edistin}
\end{figure*}

\subsection{Analysis of the  $E(b-y)$ $vs.$ distance diagrams for the  areas inside  the ring}

The same diagrams for areas ``p" to ``t", inside the ring are shown in Fig. \ref{edistin}.
The respective
coordinates are given in Table \ref{coordareas3}.

\begin{table}[htb]
\caption{ Coordinates  of  areas located inside   the ring.}
\label{coordareas3}
\begin{center}
\begin{tabular}{l|c|c|c|c}
\hline
area&$l_{min}(\degr)$&$l_{max}(\degr)$&$b_{min}(\degr)$&$b_{max}(\degr)$\\
\hline
p&330 & 350 &10 & 25 \\
q&303 & 323 &15 & 30 \\
r&290 & 310 &-18 & 2 \\
s&320 & 340 &-13 & 7 \\
t&350 & 10 &-18 & 2 \\
\hline
\end{tabular}
\end{center}
\end{table}

  The presence of the dark clouds, clearly identified by $E(b-y) \geq 0\fm100$,
can be seen on the western side at about 100 -- 150 pc and on the eastern side at about 120 -- 180 pc
in all areas, except for area ``q". Beyond these distance ranges the number
of unreddened stars decreases considerably indicating the location of these large dark cloud complexes.

 In area ``p" we have Lupus IV and Lupus V at  $d\approx$ 100 pc;
in area ``r" we have Coalsack at  $d\approx$ 160 pc and Chamaeleon at  $d\approx$ 140 pc;
in area ``t" we have RCrA at  $d\approx$ 140 pc; and in area ``s" we have Lupus I at
 $d\approx$ 150 pc.

In area ``t", near the Scutum dark cloud ($l = 25, b = 1$), there is also a first jump
to higher colour excess values at $d = 130 \pm 20$ pc and a hint of a second jump at $d = 300 \pm 40$ pc.

\subsection{Limiting magnitude effects on $E(b-y)$ and distances of the selected areas}

 To ensure that the final sample is adequate for measuring the possible existence of
an extinction jump from $E(b-y)=0.015$ to $E(b-y)= 0\fm070 \ -  \ 0\fm100$ around 100 pc we have also drawn
separate histograms of the V magnitudes,  the $\beta$ index and the absolute magnitude M$_{V}$
for each area that has been used to analyse the existence of the bubbles' interface.

\begin{figure}[htb]
\begin{center}
\includegraphics[width=0.75\linewidth]{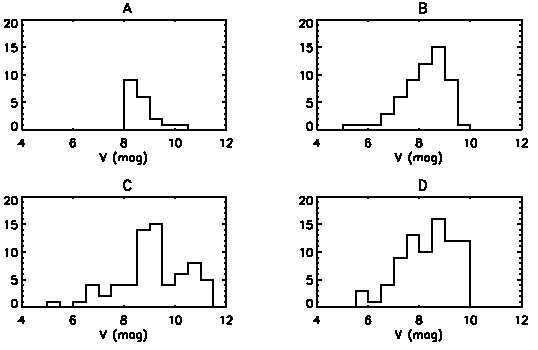}
\includegraphics[width=0.75\linewidth]{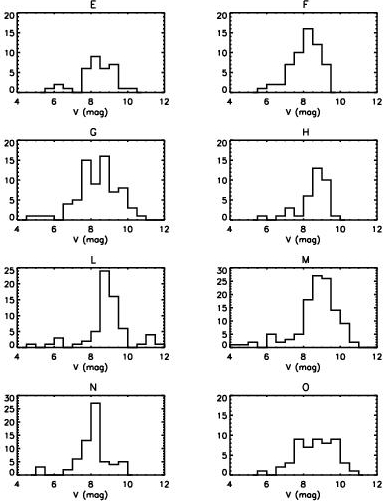}
\includegraphics[width=0.75\linewidth]{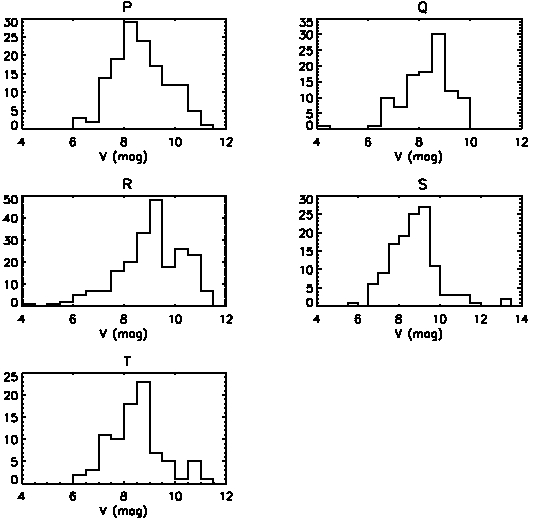}
\caption
{Histograms of the V magnitude for the analised areas.}
\label{histo_v_mag}
\end{center}
\end{figure}

 As can be seen in Fig. \ref{histo_v_mag} the histograms of the V magnitudes may be considered complete 
up to V$\approx$8.5-9.0 for all areas.

\begin{figure}[htb]
\begin{center}
\includegraphics[width=0.75\linewidth]{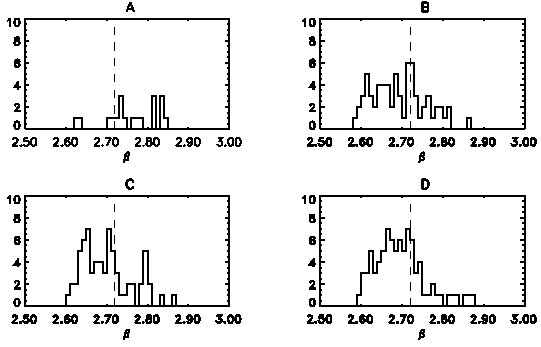}
\includegraphics[width=0.75\linewidth]{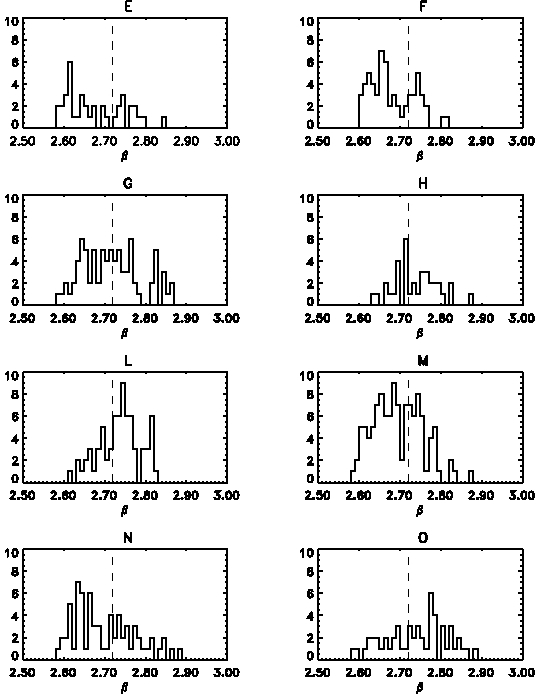}
\includegraphics[width=0.75\linewidth]{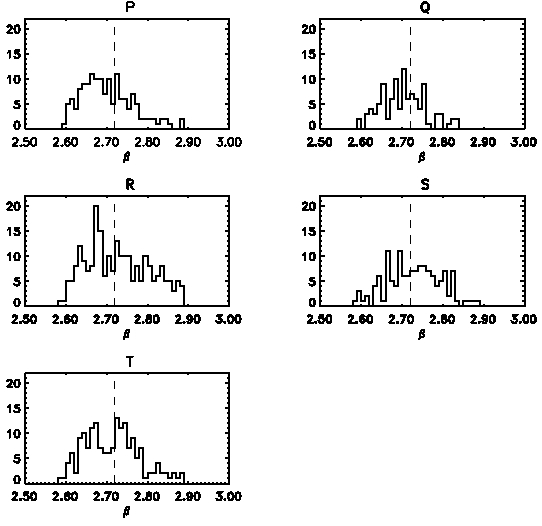}
\caption
{Histograms of the $\beta$ index for all areas. 
 The dashed line indicate $\beta$ = 2.72, which leaves the A-type stars on the right side
and the F-type stars on the left side.}
\label{histo_beta_areas}
\end{center}
\end{figure}

The histograms of the $\beta$ index, Fig. \ref{histo_beta_areas},
show that the sample has enough
A-type stars to detect the higher values of colour excess when the F-type
stars are not capable of detecting them any longer.

\begin{figure}[htb]
\begin{center}
\includegraphics[width=0.75\linewidth]{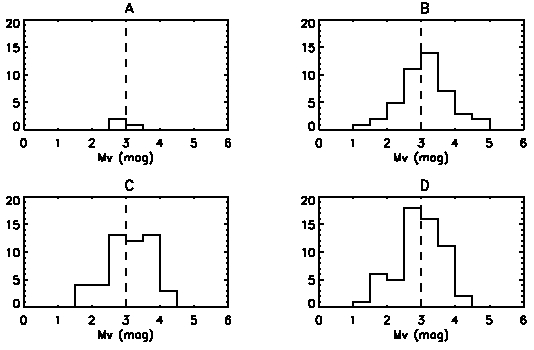}
\includegraphics[width=0.75\linewidth]{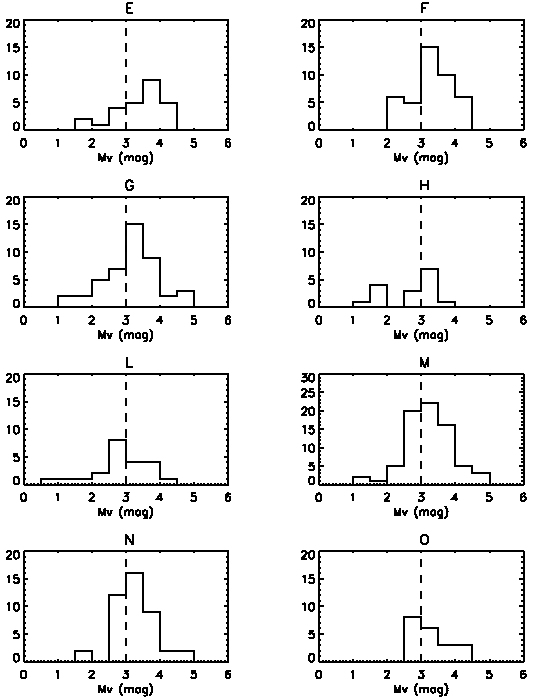}
\includegraphics[width=0.75\linewidth]{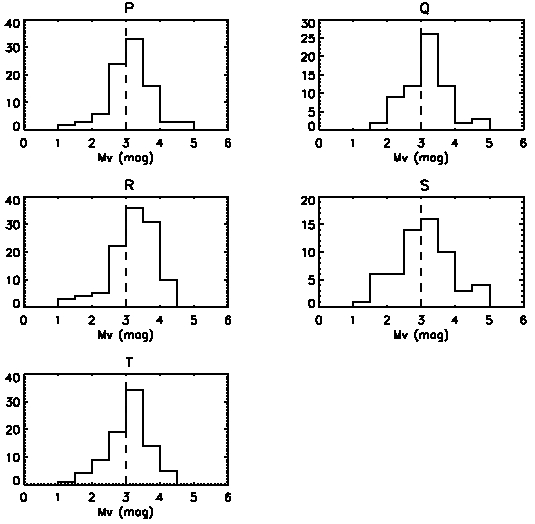}
\caption
{Histograms of the M$_{V}$ for F-type stars for all analised areas. The dashed line indicates M$_{V}=3\fm0$ }
\label{histo_mv_f_areas}
\end{center}
\end{figure}

\begin{figure}[htb]
\begin{center}
\includegraphics[width=0.75\linewidth]{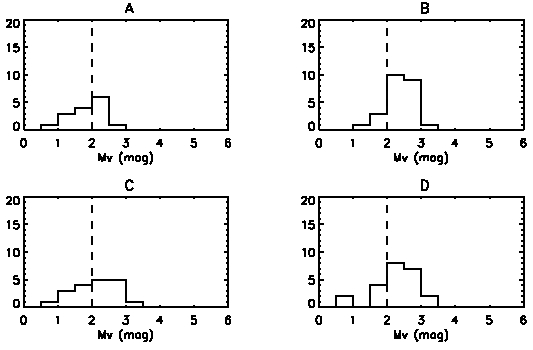}
\includegraphics[width=0.75\linewidth]{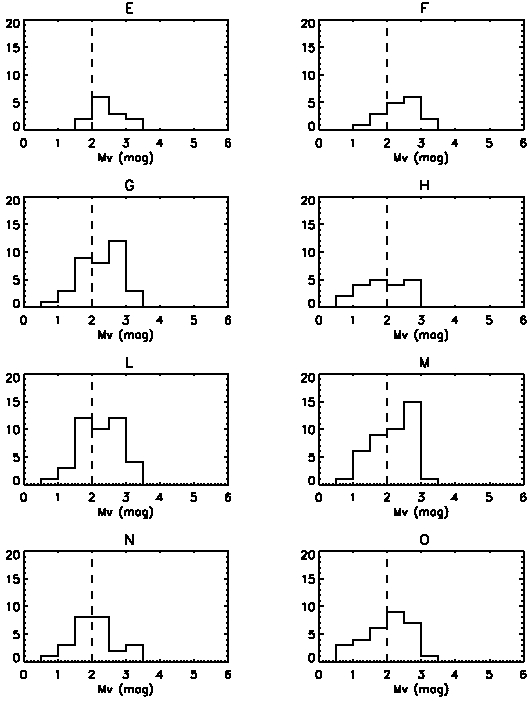}
\includegraphics[width=0.75\linewidth]{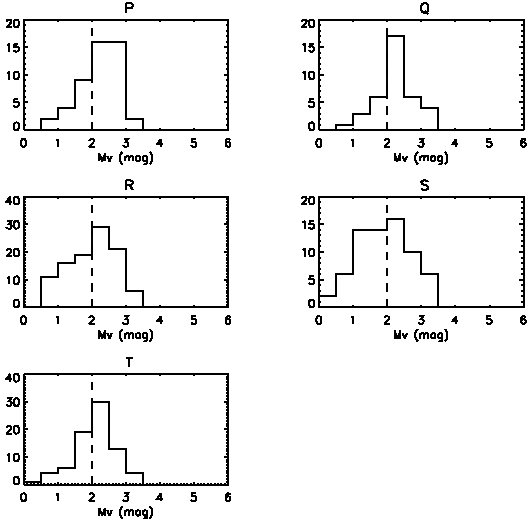}
\caption
{Histograms of the M$_{V}$ for A-type stars for all analised areas. The dashed line indicates M$_{V}=2\fm0$ }
\label{histo_mv_a_areas}
\end{center}
\end{figure}

As can be inferred from Fig. \ref{histo_mv_f_areas} and \ref{histo_mv_a_areas},
the typical M$_{V}$ in the selected areas is around $3\fm0$ and $2\fm0$ for
the F- and A-type stars, respectively. 

Taking one of the lowest limiting magnitude cases among the studied areas, V = $8\fm5$,
the F-type stars of the sample could have detected $E(b-y) = 0\fm100$ up to 103 pc and the
A-type stars up to a distance of 164 pc. If we consider $E(b-y) = 0\fm050$, the 
corresponding distances would be 114 pc and 180 pc for the F- and A-type stars, respectively.
Thus, our sample is able to detect a jump in colour excess from $0\fm015$ to $0\fm100$ 
around 100 pc.

 Looking at the histograms of $\beta$ and M$_{V}$  -- Figs. \ref{histo_beta_areas},
\ref{histo_mv_f_areas} and \ref{histo_mv_a_areas} -- one might infer that there are no selection
effects on the colour excess $vs.$ distance diagrams. There are regions with higher limiting magnitude
and better spectral type coverage where greater reddenings are not detected, whereas there are 
areas with worse spectral coverage and lower limiting magnitude where greater reddenings are detected.
Therefore, we conclude that if the reddening were present it could be detected.

\section{Discussion }

Analyzing these diagrams, we can summarize the results in the following way:
\begin{itemize}

\item Until 60 - 80 pc the colour excess remains below $E(b-y) \leq 0\fm040$ in all directions,
with $0\fm020$ a being typical value.

\item The expected transition to $E(b-y) \approx  0\fm070 - 0\fm100$ occurs on the western  part of the ring
at $d=110 \pm$ 20 pc, whereas on the eastern side it is not clearly seen before 280 $\pm$ 50 pc.

\item In the northernmost area of the ring there seems to be a gradual increase from $0\fm020$ to around
$0\fm050$, while in the southernmost area the colour excess remains constant up to the maximum surveyed distance.
However, a possible transition is observed at 200 $\pm$ 20 pc towards the Mensa constellation in this region.

\item In the internal part of the ring the main features are defined by the dark clouds, clearly identified by
$E(b-y) \geq 0\fm100$ on the western side at about 100 -- 150 pc and on the eastern side at about 120 -- 180 pc. 
Beyond these distance ranges the number of unreddened stars decreases considerably indicating the location of
these large dark cloud complexes.

\item In the external parts of the ring the colour excess remains below $0\fm040$ in all directions.

\end{itemize}

Comparing our results with the literature ones  we observe that \citet{sfeir99}
identify a tunnel in  $l=330\degr$, $b\approx 12\degr$ until $d\approx 160$ pc. 
The existence of this tunnel is confirmed in our work, since  in this direction
we find  $E(b-y) \leq 0\fm010$ until $d \approx 150$ pc. \citet{perry} suggests that the effects of
interstellar reddening may be ignored for stars within 80 pc in agreement with our results.

On the other hand, \citet{centurion91} analyzing UV
spectra  of eight stars in the region defined by the Galactic coordinates
$310\degr \leq l \leq 330\degr$ and $15\degr \leq b \leq 25\degr$ observe 
an increase in the column density of the neutral sodium at $d = 40 \pm 25$ 
pc and suggest that this is the distance of the interface between the LB 
and  Loop I. However, in our work, we verify that up to 60 pc 
$E(b-y) \leq 0\fm020$ in all directions and that in this direction, in 
particular, we have stars considered without  reddening ($E(b-y) \leq 0\fm010$) 
up to  180 pc (see Fig. \ref{060}).

\citet{knude98} estimated the distance to the star forming clouds in the Lupus IV and Lupus V region as around
100 pc. \citet{franco02} report a somewhat greater distance of 150 pc to Lupus I. We also observe a similar
behaviour in the Lupus direction, which suggests that the Lupus region may have two distinct structures.

Our results for the infrared filament towards the Mensa constellation (210 $\pm$ 20 pc) supports the distance
determination of 230 $\pm$ 30 pc by \citet{penprase98}. This reinforces our findings that the eastern and
western parts of the ring are located at  different distances.

 \citet{frisch2007} uses mean extinction maps to study the distribution of the ISM surrounding the LB.
$E(B-V)$ is calculated from the photometric data in the Hipparcos catalog and averaged over stars in a sector
with a width of $\pm 13\degr$ and overlapping distance uncertainties.
 Like us, \citet{frisch2007} sees the same large scale features that would compose  the ring, and a similar
spatial distribution of the colour excess. Since her map samples the whole distance interval up to 500 pc it
is difficult to infer the precise location of the  material, but the mean extinction values are also very
different to both the eastern and western sides of the ring.

 The same difference of the gas column density between the eastern and western sides of the ring
can be seen in Fig. 3c of \citet{park07} that shows a N$_{H}$ map for $b \geq 25\degr$ and
$90\degr \leq l \leq -90\degr$. In such map N$_{H}$ changes from
$\approx 7 \times 10^{20} cm^{-2}$ $(E(b-y) \approx 0\fm100)$ on the western side to
$\approx 4.5 \times 10^{20} cm^{-2}$ $(E(b-y) \approx 0\fm070)$ on the eastern side.

 \citet{wolleben2007} calls attention to the fact that the sky projection of his two overlapping
synchroton shells, although resembling a ring-like structure, are in fact the result of two super-imposed
H {\sc I} shells that are expanding with different velocities.

Taken together these results suggest that either the ring is not a real entity or it is
very fragmented and highly distorted.


\section{Conclusions}

In the reddening distribution analysis we have tried to identify the ring suggested 
by \citet{egger95}, since it would be the most prominent feature of the interaction region.
The results can be summed up as follows:

\begin{itemize}
\item Until 60 - 80 pc the colour excess remains below $E(b-y) \leq 0\fm040$ in all directions,
$0\fm020$ being a typical value.

\item The expected transition to $E(b-y) \approx  0\fm070 - 0\fm100$ occurs on the western
 part of the ring about 110$\pm$ 20  pc,  whereas on the eastern
side it is not clearly seen before 280 $\pm$ 50 pc. 

\item In the northernmost area of the ring there seems to be a gradual increase from $0\fm020$ to around
$0\fm050$, while in the southernmost area the colour excess remains constant up to the maximum surveyed distance.
However, a possible transition is observed at 200 $\pm$ 20 pc towards the Mensa constellation in this region.

\item In the internal part of the ring the main features are defined by the dark clouds, clearly identified by
$E(b-y) \geq 0\fm100$. Our results corroborate their earlier distance determination, that is Lupus IV and Lupus V
at $d = 110 \pm 20$ pc; Lupus I and Oph-Sgr at $ 140 \pm 20$ pc; Aql-South, Sag-South and RCrA at $d = 140 \pm 20$ pc;
Coalsack $d= 160 \pm 20$ pc; Chamaeleon and G317-4 at $d= 140 \pm 20$ pc; Scutum with two main parts at $d=130 \pm 20$ pc
and $d=300 \pm 40$ pc, respectively.  
\end{itemize}

If the ring-like feature, as proposed by \citet{egger95}, really exists the colour excess distribution suggests
that it is  very fragmented and extremely distorted, once its western side reaches at $d= 110 \pm 20$ pc
and its eastern side  at $d =280 \pm 50$ pc. However, the very different characteristics of the reddening inside 
and along the ring-like feature do not support the hypothesis of a ring.

\begin{acknowledgements}
 W. Reis and W.J.B. Corradi acknowledge support from the Brazilian Agencies CAPES and
 Fapemig (Grants No. EDT 1883/03 and CEX 961/04).
\end{acknowledgements}

\bibliographystyle{aa}

\end{document}